\documentclass[12pt,a4paper]{article}

\usepackage[utf8]{inputenc} 
\usepackage[T1]{fontenc} 
\usepackage[english]{babel} 
\usepackage{graphicx} 
\usepackage{amsmath} 
\usepackage{geometry} 
\usepackage{booktabs} 
\usepackage{array} 
\usepackage[hyphens]{url} 
\usepackage{hyperref} 
\usepackage{authblk}
\usepackage{microtype} 

\hypersetup{
    colorlinks=true,
    linkcolor=blue,
    filecolor=magenta,
    urlcolor=cyan,
    citecolor=red,
}

\newcolumntype{L}[1]{>{\raggedright\arraybackslash}p{#1}}
\newcolumntype{R}[1]{>{\raggedleft\arraybackslash}p{#1}}
\newcolumntype{C}[1]{>{\centering\arraybackslash}p{#1}}

\title{Case Study: Performance Analysis of a Virtualized XRootD Frontend in Large-Scale WAN Transfers}
\author[1]{J M da Silva}
\author[1]{M A Costa}
\author[1]{R L Iope}

\affil[1]{Núcleo de Computação Científica, São Paulo State University (UNESP), São Paulo, Brazil}

\date{\today}

\begin{document}

\maketitle

\begin{abstract}
This paper presents a detailed case study of the T2\_BR\_SPRACE storage frontend architecture and its observed performance in high-intensity data transfers. The architecture is composed of a heterogeneous cluster of XRootD \cite{xrootd} Virtual Machines (VMs) with 10 Gb/s and 40 Gb/s links, which aggregate data from a 77 Gb/s dCache \cite{dcache} backend via pNFS to an external 100 Gb/s WAN link. We describe the system configuration, including the use of the BBR \cite{bbr} congestion control algorithm and TCP extensions \cite{rfc1323}. Under peak production conditions, we observed the system sustaining an aggregate throughput of 51.3 Gb/s. An analysis of a specific data flow to Fermilab (FNAL) showed peaks of 41.5 Gb/s, validated by external monitoring tools (CERN). This study documents the performance of a complex virtualized architecture under real load.
\end{abstract}

\section{Introduction}

Distributed computing for High Energy Physics experiments, such as CMS at the LHC \cite{cms}, depends on the efficient transfer of petabytes of data across the Worldwide Computing Grid (WLCG) \cite{wlcg}. A common challenge for computing centers (Tiers) is to aggregate heterogeneous storage endpoints to efficiently saturate high-capacity WAN (Wide Area Network) links.

This article presents a case study of the T2\_BR\_SPRACE storage endpoint. Unlike an optimization study focused on proving causality, this work focuses on describing the production architecture and documenting the observed performance during a period of high demand. The SPRACE architecture uses a frontend of XRootD \cite{xrootd} Virtual Machines (VMs) with 10 Gb/s and 40 Gb/s connectivity (some with SR-IOV) to aggregate data from a high-performance dCache \cite{dcache} backend.

The objective of this study is to document this heterogeneous virtualized architecture in detail (Sections 2.1, 2.2, 2.3) and to present the observed peak throughput results (51.3 Gb/s aggregate and 41.5 Gb/s in a WAN flow) under this configuration (Section 3).

\section{Architecture and Configuration Description}

The observed performance is the result of the complex interaction between the backend, frontend components, and the applied network configuration.

\subsection{System Architecture}

\begin{figure}[hbt!]
\centering
\includegraphics[width=0.8\textwidth]{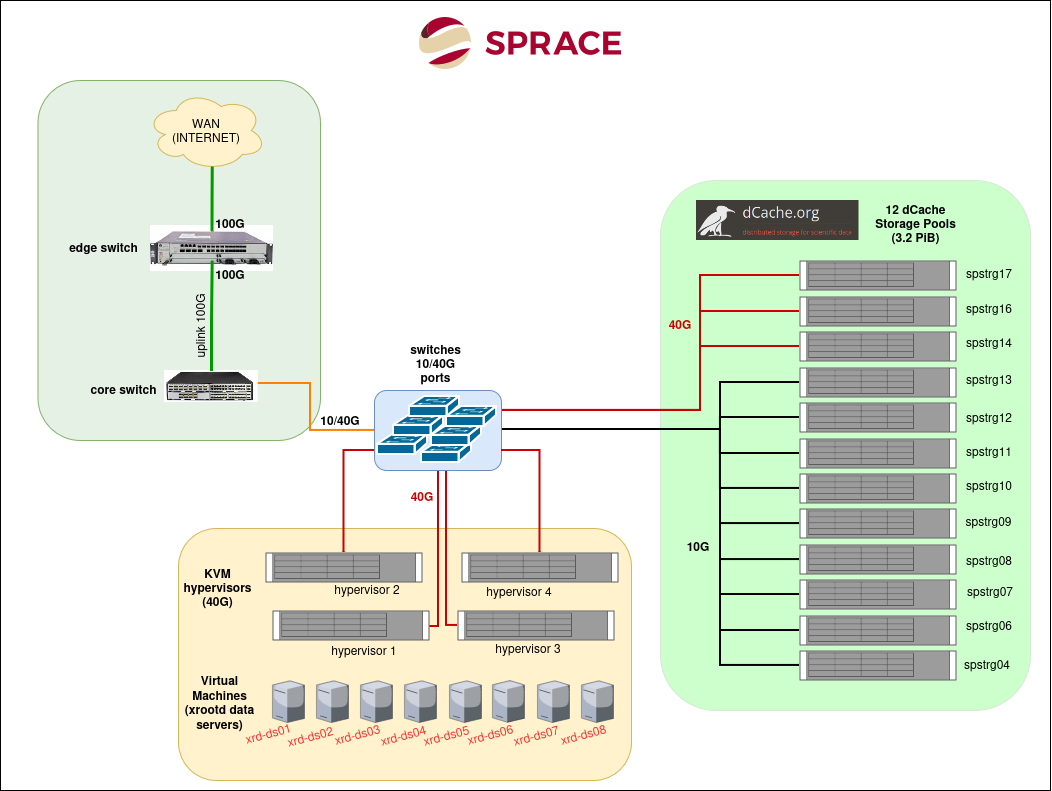}
\caption{Storage cluster architecture.}
\label{fig:figura4}
\end{figure}

The analyzed system is a storage cluster with a decoupled architecture, Figure \ref{fig:figura4}:
\begin{itemize}
    \item \textbf{Backend:} The data layer consists of 12 dCache \cite{dcache} storage pools. These serve data to the frontend servers via the pNFS protocol (parallel NFS 4.1), eliminating a single NFS gateway bottleneck.
    \item \textbf{Frontend (Data Access):} The frontend is composed of 8 XRootD \cite{xrootd} servers operating as Virtual Machines (VMs). These servers are the endpoints that handle data transfers to the WAN.
    \item \textbf{Virtualization and Network:} The VMs run on physical hypervisors equipped with 10 Gb/s and 40 Gb/s NICs. The VMs on the 40 Gb/s hosts use SR-IOV for direct access (PCI passthrough) to the network hardware.
    \item \textbf{Management:} Traffic is managed by 1 local XRootD redirector and transfers are orchestrated by the CERN FTS (File Transfer Service) \cite{fts}.
\end{itemize}

\subsection{\textit{Frontend} Configuration (XRootD Servers)}

The \textit{frontend} VM \textit{cluster} is heterogeneous in terms of allocated memory (varying from approximately 11.4 GiB to 15.4 GiB). Table \ref{tab:vm_config} details the memory configuration of each VM and the applied TCP \textit{buffer} ceilings.

\begin{table}[hbt!]
\centering
\caption{Memory specifications (allocated to VMs) and TCP buffer on the 8 XRootD servers.}
\label{tab:vm_config}
\scriptsize 
\begin{tabular}{@{}lccccccl@{}}
\toprule
\textbf{Server} & \textbf{Cores} & \textbf{Memory} & \textbf{Buffers} & \textbf{Used for TCP} & \textbf{Max. Recv} & \textbf{Max. Send} & \textbf{TCP Memory} \\
\textbf{(VM)} & & \textbf{(MiB)} & \textbf{XRootD (MiB)} & \textbf{Buffers (MiB)} & \textbf{Buffer (MiB)} & \textbf{Buffer (MiB)} & \textbf{Pages} \\
\midrule
xrd-ds01 & 12 & 11702 & 2925 & 6600 & 2048 & 2048 & 1689848 \\
xrd-ds02 & 12 & 15734 & 3933 & 7704 & 2048 & 2048 & 1972471 \\
xrd-ds03 & 12 & 15734 & 3933 & 7704 & 2048 & 2048 & 1972472 \\
xrd-ds04 & 12 & 11702 & 2925 & 4680 & 2048 & 2048 & 1198329 \\
xrd-ds05 & 12 & 15734 & 3933 & 7704 & 2048 & 2048 & 1972472 \\
xrd-ds06 & 12 & 15734 & 3933 & 7704 & 2048 & 2048 & 1972472 \\
xrd-ds07 & 12 & 15732 & 3933 & 7703 & 2048 & 2048 & 1972094 \\
xrd-ds08 & 12 & 11702 & 2925 & 4680 & 2048 & 2048 & 1198330 \\
\bottomrule
\end{tabular}
\end{table}

The observed performance directly depends on a set of kernel-level optimizations. The default operating system settings, usually focused on the CUBIC \cite{cubic} algorithm and limited TCP memory buffers, are known to be insufficient for saturating high-latency, high-bandwidth WAN links (like the 100 Gb/s one) \cite{esnet}.

\begin{sloppypar}
The set of applied kernel-level optimizations are detailed in \texttt{99-xrootd-tcp-tunnings.conf}, whose key parameters include:

\begin{itemize}
    \item \textbf{BBR Congestion Control:} Google's BBR (Bottleneck Bandwidth and Round-trip propagation time) algorithm \cite{bbr} was enabled (\texttt{net.ipv4.tcp\_congestion\_control=bbr}).
    \item \textbf{Network Buffers (TCP Buffers):} As shown in Table \ref{tab:vm_config}, the maximum memory allocation limits (\texttt{net.core.rmem\_max} and \texttt{net.core.wmem\_max}) were expanded to a ceiling of 2048 MiB (2147483647 bytes).
    \item \textbf{Window Scaling and Timestamps:} The \texttt{net.ipv4.tcp\_window\_scaling = 1} and \texttt{net.ipv4.tcp\_timestamps = 1} optimizations were enabled, as defined in RFC 1323 \cite{rfc1323}, to allow the receive window to exceed the 64 KB limit.
    \item \textbf{Advertised Window Scale:} \texttt{net.ipv4.tcp\_adv\_win\_scale = 1} was set. This parameter instructs the kernel to use a fraction of the buffer space (specifically, 1/2\textsuperscript{1}, or 50\%) as the advertised receive window. This means that, although the memory allocation ceiling is 2048 MiB, the buffer effectively advertised to the sender is at most 1024 MiB, creating a safety margin for the application (XRootD) to read data from the socket.
    \item \textbf{Connection Optimizations:} The limits for pending connections (\texttt{net.core.somaxconn = 65536}) and the network device backlog (\texttt{net.core.netdev\_max\_backlog = 65536}) were increased.
\end{itemize}
\end{sloppypar}

It is crucial to note the synergy between these settings. BBR \cite{bbr} is the algorithm that determines the optimal sending rate. However, to execute this decision, it depends on the kernel's buffer auto-tuning mechanism to dynamically allocate the necessary memory (SO\_SNDBUF) for each TCP flow. This dynamic auto-tuning is only effective because the maximum memory allocation limit (the 2048 MiB "ceiling") was set manually, and the advertised window (controlled by \texttt{tcp\_adv\_win\_scale}) was limited to 1024 MiB, allowing the kernel to safely manage massive buffers, far beyond what OS defaults would permit.

\subsection{\textit{Backend} Configuration (Disk I/O)}

The I/O capacity of the \textit{backend} (the 12 dCache \textit{pools}) was measured individually. Table \ref{tab:io_backend} presents the results of direct read tests (copying 1 GiB of data) from each \textit{pool}.

\begin{table}[hbt!]
\centering
\caption{Disk I/O (read) tests on the 12 dCache pools.}
\label{tab:io_backend}
\begin{tabular}{@{}lcc@{}}
\toprule
\textbf{Pool Host} & \textbf{Time (s)} & \textbf{Speed (MB/s)} \\
\midrule
spstrg04 & 1.66071 & 647 MB/s \\
spstrg06 & 1.64576 & 652 MB/s \\
spstrg07 & 1.63163 & 658 MB/s \\
spstrg08 & 1.65356 & 649 MB/s \\
spstrg09 & 1.46332 & 734 MB/s \\
spstrg10 & 1.39045 & 772 MB/s \\
spstrg11 & 1.38576 & 775 MB/s \\
spstrg12 & 1.42805 & 752 MB/s \\
spstrg13 & 1.46831 & 731 MB/s \\
spstrg14 & 0.888589 & 1200 MB/s* \\
spstrg16 & 1.05212 & 1000 MB/s* \\
spstrg17 & 1.00796 & 1100 MB/s* \\
\midrule
\textbf{TOTAL} & & \textbf{9,670 MB/s} \\
\bottomrule
\end{tabular}
\footnotesize{\\ * These pools are hosted on newer generation servers equipped with improved disk controllers.}
\end{table}

The maximum theoretical aggregate I/O throughput of the backend is 9,670 MB/s, which is equivalent to 77.36 Gb/s.

\section{Observed Results}

The analysis focuses on a period of high transfer demand in the early morning of October 17, 2025.

\subsection{Aggregate Performance}

\begin{figure}[hbt!]
\centering
\includegraphics[width=0.8\textwidth]{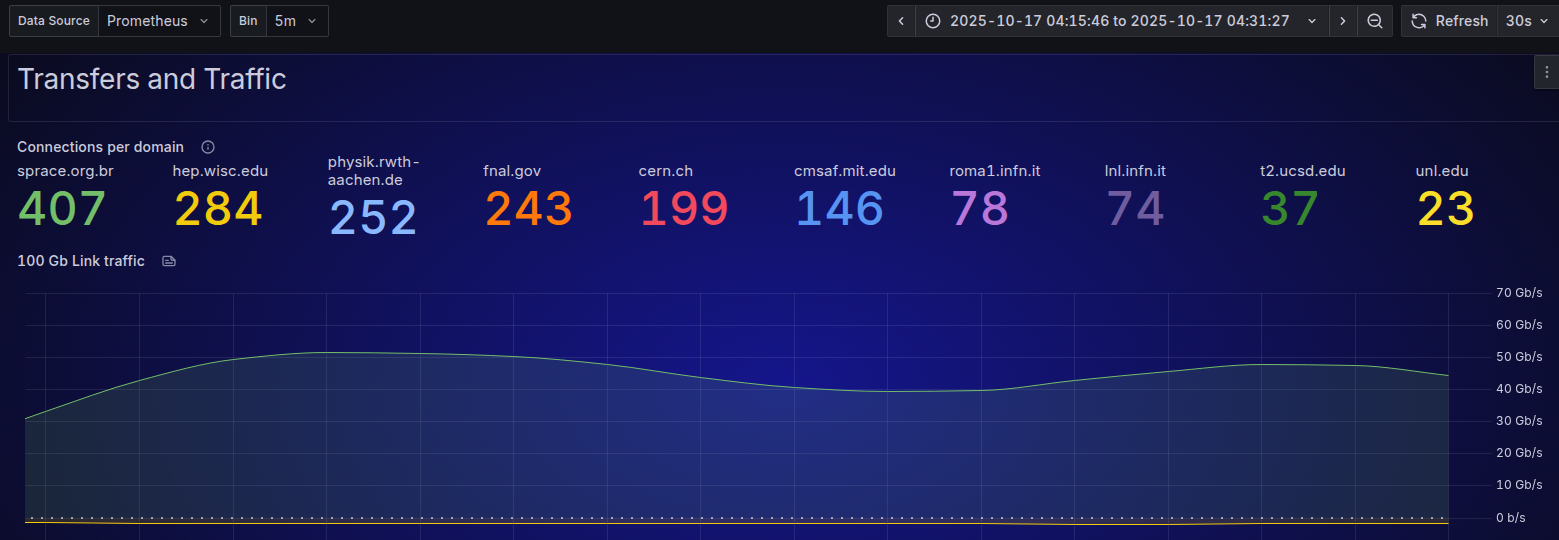}
\caption{Aggregate throughput.}
\label{fig:figura1}
\end{figure}

During the monitored period, the system reached a peak aggregate throughput of 51.3 Gb/s, as illustrated in Figure \ref{fig:figura1}. This value represents the total outbound traffic from the XRootD cluster, aggregated from the 8 VMs, to the WAN.

\begin{figure}[hbt!]
\centering
\includegraphics[width=0.8\textwidth]{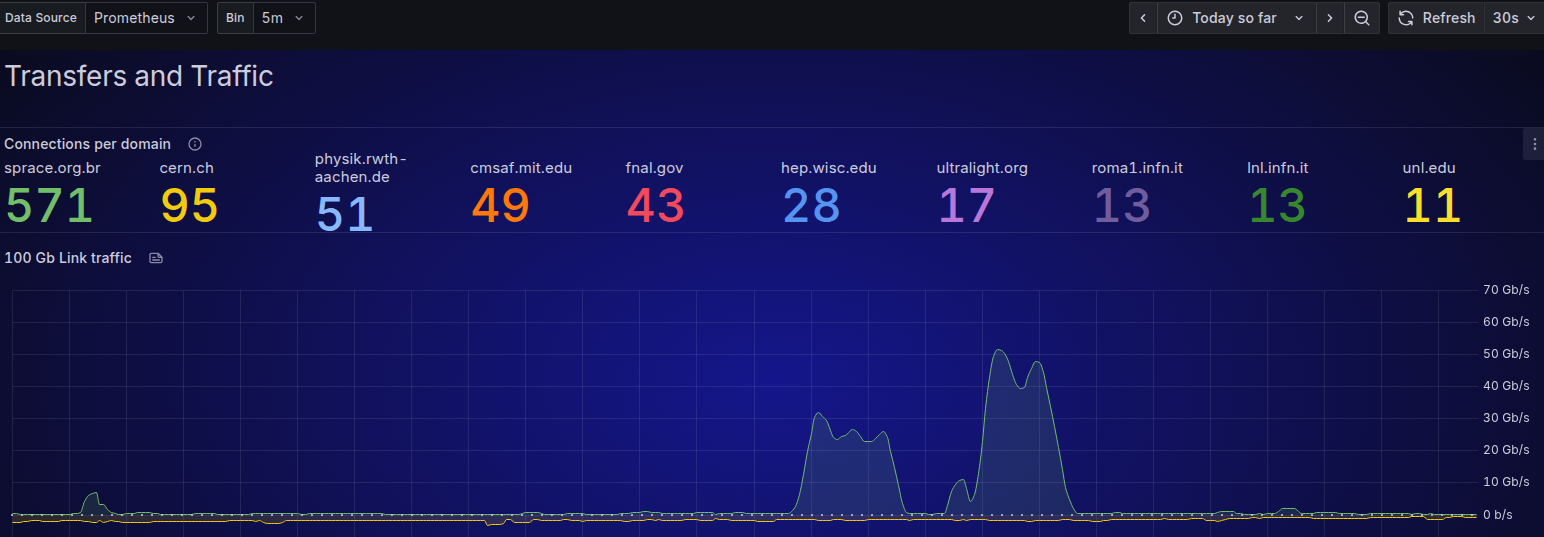}
\caption{Connection distribution.}
\label{fig:figura2}
\end{figure}

The system processed a total of 5,696 completed transfers, of which 4,443 (78.00\%) were successful. The failures (22.00\%) are discussed in Section 4.1. Figures \ref{fig:figura1} and \ref{fig:figura2} show the distribution of active connections, highlighting the predominance of local transfers (sprace.org.br) and to large centers like FNAL, CERN, and Wisconsin.

\subsection{Peak WAN Flow Analysis (SPRACE $\rightarrow$ FNAL)}

Table \ref{tab:fluxo_fnal} details the temporal evolution of the throughput on the \url{davs://osg-se.sprace.org.br} $\rightarrow$ \url{davs://cmsdcadisk.fnal.gov} link, which represented the main WAN data flow during the peak period.

\begin{table}[hbt!]
\centering
\caption{Throughput evolution on the SPRACE $\rightarrow$ FNAL link during the peak period (04:04 - 04:32 BRT).}
\label{tab:fluxo_fnal}
\begin{tabular}{@{}lccccc@{}}
\toprule
\textbf{Time} & \textbf{Active} & \textbf{Queue} & \textbf{Success} & \textbf{Throughput} & \textbf{Streams} \\
\textbf{(BRT)} & & & \textbf{Rate} & & \textbf{(Decision)} \\
\midrule
04:04:33 & 56 & & 100\% & 24.16 Mb/s & 300 \\
04:06:16 & 0 & & 100\% & 1.51 Gb/s & 300 \\
04:12:03 & 347 & & 100\% & 2.96 Gb/s & 310 (+10) \\
04:13:12 & 538 & & 100\% & 5.24 Gb/s & 330 (+10) \\
04:16:02 & 737 & & 100\% & 13.66 Gb/s & 380 (+10) \\
04:22:47 & 764 & & 100\% & 29.29 Gb/s & 500 (+10) \\
04:24:28 & 527 & & 100\% & 33.07 Gb/s & 530 (+10) \\
04:27:50 & 193 & & 100\% & 41.48 Gb/s & 530 \\
04:32:24 & 45 & 2 & 100\% & 36.59 Gb/s & 530 \\
\bottomrule
\end{tabular}
\end{table}

During this 28-minute period, the FTS \cite{fts} scaled the number of streams from 300 to 530. The aggregate throughput on this specific link reached a peak of 41.5 Gb/s (41.48 Gb/s) and, notably, no failures were recorded on this link during this interval.

\begin{figure}[hbt!]
\centering
\includegraphics[width=0.8\textwidth]{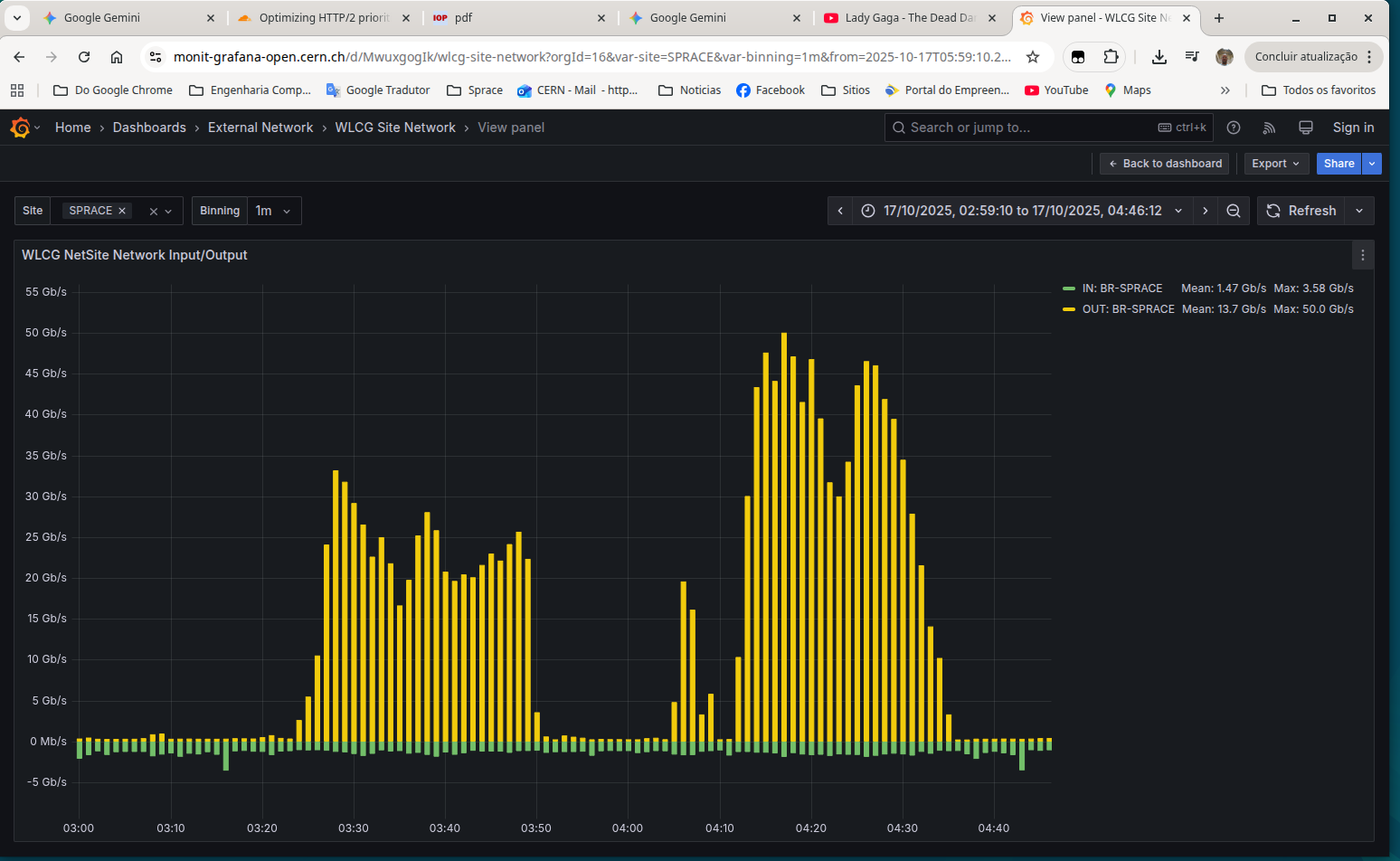}
\caption{CERN (FTS) monitoring validating the throughput peaks above 40 Gb/s in the analyzed period.}
\label{fig:figura3}
\end{figure}

These local measurements are corroborated by independent monitoring tools. Figure \ref{fig:figura3} shows the view from the CERN monitoring for the same period, recording throughput peaks exceeding 40 Gb/s, aligned with the 41.5 Gb/s measured locally.

\section{Discussion}

\subsection{Bottleneck Analysis}

The maximum backend performance (77.36 Gb/s, Table \ref{tab:io_backend}) is significantly higher than the measured peak aggregate traffic (51.3 Gb/s). This indicates that disk I/O was not the limiting bottleneck.

The bottleneck seems to lie in the frontend's capacity. The 51.3 Gb/s limit is likely the result of the combined saturation of the 8 VMs' resources (10G/40G NIC limits, CPU contention, or RAM exhaustion for TCP buffers, as per Table \ref{tab:vm_config}) while managing 530 concurrent BBR streams, in addition to the pNFS backend I/O.

\subsection{Reliability Analysis}

The overall success rate of 78.00\% (Section 3.1) is explained by the concentration of failures on specific destination endpoints. Notably, the \url{cmstnvm1.fnal.gov} destination showed a 100\% failure rate for all 50 source sites, indicating a problem with the destination service.

In contrast, the main data destinations showed high reliability in the overall period:
\begin{itemize}
    \item \url{cmsdcadisk.fnal.gov}: 97.92\% success (2,635 OK, 56 failures).
    \item \url{eoscms.cern.ch}: 97.93\% success (1,375 OK, 29 failures).
\end{itemize}

The observation of zero failures on the FNAL link (Table \ref{tab:fluxo_fnal}) during the maximum stress of 41.5 Gb/s suggests that the local configuration (VMs + BBR) was stable under high load.

\section{Conclusion}

This case study documented the architecture and peak performance of a virtualized and heterogeneous storage frontend. We demonstrated that a cluster of 8 XRootD VMs (with 10G/40G NICs and SR-IOV) was capable of aggregating traffic from a dCache/pNFS backend (77 Gb/s) and sustaining a WAN output throughput of 51.3 Gb/s. The limiting bottleneck was not disk I/O, but rather the aggregate capacity of the frontend VM cluster (RAM, CPU, or NIC limits).

Notably, a single data flow to FNAL reached 41.5 Gb/s, validated by external monitoring, demonstrating the setup's ability to operate successfully under extreme load. This result validates that the applied network configuration, far superior to OS defaults for this purpose, (including BBR tuning and FTS stream scaling) was a critical component in enabling the architecture to achieve this fraction of the 100 Gb/s link.

\bibliographystyle{ieeetr} 
\bibliography{artigo-en}  

\end{document}